\documentclass[prb,showpacs,twocolumn]{revtex4}
\usepackage{graphicx}
\usepackage{amsmath}
\begin{document}

\let\veps=\varepsilon%
\newcommand\tr{\mathop{\mathrm{Tr}}}%
\newcommand\half{\frac{1}{2}}%
\newcommand\lavg{\left\langle}%
\newcommand\ravg{\right\rangle}%
\newcommand\varE{{\mathcal E}}%
\newcommand\im{\mathop{\mathrm{Im}}\nolimits}%
\newcommand\ket[1]{\left|#1\right\rangle}%
\newcommand\bra[1]{\left\langle#1\right|}%
\newcommand\braket[2]{\left.\left\langle#1\right|#2\right\rangle}

\title{Charge and current fluctuations in a superconducting single
electron transistor near a Cooper pair resonance}

\author{Mahn-Soo Choi}

\affiliation{Korea Institute for Advanced Study, Cheongryanri-dong
207-43, Seoul 130-012, Korea}

\author{Francesco Plastina and Rosario Fazio}

\affiliation{NEST-INFM $\&$ Scuola Normale Superiore, I-56127
Pisa,Italy}

\author{cond-mat/0208165}

\date{\today}

\begin{abstract}
We analyze charge tunneling statistics and current noise in a
superconducting single-electron transistor in a regime where the
Josephson-quasiparticle cycle is the dominant mechanism of
transport. Due to the interplay between Coulomb blockade and
Josephson coherence, the probability distribution for tunneling
events strongly deviates from a Poissonian and displays a
pronounced even--odd asymmetry in the number of transmitted
charges. The interplay between charging and coherence is reflected
also in  the zero-frequency current noise which is significantly
quenched when the quasi-particle tunneling rates are comparable to
the coherent Cooper-pair oscillation frequency. Furthermore the
finite frequency spectrum shows a strong enhancement near the
resonant transition frequency for Josephson tunneling.
\end{abstract}

\maketitle

\section{Introduction}
\label{long:sec1}

Shot noise in a mesoscopic conductor is a consequence of the
stochastic character of electron tunneling and of the discreteness of
charge.  Unlike thermal noise, shot noise describes the
non-equilibrium fluctuations of current; therefore, the study of
current fluctuations can provide further understanding of properties
related to correlation mechanisms, internal energy scales or the carrier
statistics which cannot be obtained by measuring the average
current~\cite{deJong97,Blanter00,Davies92,Hershfield93}.

A well studied example of physical processes where electron
correlations play a dominant role is the phenomenon of Coulomb
blockade. In a system of small tunneling junctions, due to the
large electrostatic energy (as compared to temperature or
voltages), the electronic charge is transported one by one.  This
effect leads to many remarkable features in transport properties
and has been a subject of extensive study for the last
decades~\cite{Schoen90,Averin91}. As an example, the strong
dependence of the current-voltage characteristics on the gate
charge was exploited to use a single electron transistor (SET) as
a highly sensitive charge detector~\cite{Schoelkopf98}, and
proposed as measuring apparatus of the charge state of a Josephson
quantum bit~\cite{Makhlin00a,Devoret00,Averin00,chalm}.  Since it
leads to a strong correlation of consecutive tunneling events,
Coulomb blockade has turned out to manifest itself in a peculiar
way on the current noise. Such an effect has been studied both in
the sequential
tunneling~\cite{Davies92,Hershfield93,Hanke93-94,Korotkov94a} and
in the cotunneling regime~\cite{Sukhorukov00,Averin00a}.
Additional interest in studying noise in single electron devices
comes from the recently proposed schemes that employ current
fluctuation measurements to detect entanglement in solid state
systems~\cite{Loss00}.

An even richer scenario occurs when the coherence of charge carriers
is maintained over a significant portions of the system.  Such a
circumstance is encountered quite often when the tunneling junctions
are superconducting.  In this case, the charge carriers are Cooper
pairs, and their coherent tunneling across the junctions gives rise
to a series of pronounced structures in the $I$-$V$ characteristics at
sub-gap voltages~\cite{Fulton89a,Averin89,vandenBrink91}.
Furthermore, as analyzed in Ref.~\onlinecite{Averin96a}, the scattering of
quasi-particles (and consequently the shot noise) in superconducting point
contacts is significantly enhanced in the presence of the supercurrent
produced by a coherent flow of Cooper pairs.

In this paper, we analyze a superconducting double tunnel junction
device, operating in a suitably chosen bias voltage regime, such that
one of the junctions of the SET is on resonance for Cooper pair tunneling
(the case where Cooper pair resonance occurs on both junctions has been
recently analyzed in Ref.~\onlinecite{Clerk02}). The interplay between
coherence and interaction is explored by sweeping the operating point of the
device through the Cooper pair resonance.  We will show that
the fluctuations of the charge on the central island are
sensitive to both Coulomb blockade and quantum coherence.  More
pronounced effects arise in the regime in which the rates of
incoherent quasi-particle tunneling matches the frequency of coherent
Cooper-pair oscillation.  This gives rise to an enhanced fluctuation
of charge in the central-island and to a substantial suppression of the
current noise.  By investigating the statistics of the tunneling
events, we show that the suppression in the shot noise is related to
the deviation of the counting statistics from the Poissonian
distribution.  The probability distribution of tunneling events
exhibits a parity dependence and remains non-Poissonian in a wide
range of parameter values.  The interplay between coherence and
Coulomb blockade affects the overall charge transport and is also
clearly observed in the finite frequency behavior of the current
noise. Its power spectrum displays a sharp resonance peak at the
Josephson frequency, resulting from coherent oscillations between two
quantum states.

The work presented here applies to the setup used in a recent
experiment~\cite{Nakamura99} to probe the coherent
evolution of quantum states in a Cooper pair box as well as in an
earlier experiment~\cite{Fulton89a} on resonant
Cooper pair tunneling.  In this paper we extend the results of
Ref.~\onlinecite{ChoiMS01c}.  The paper is organized as follows.
In Section~\ref{long:sec2}, we introduce the model to describe the SET
transistor and we describe the relevant processes involved in the
Josephson-quasiparticle cycle. In the same section we also introduce the
master equation, in its general form, used to obtain all the results of
this work. The solution of the master equation is worked out to analyze
in detail the fluctuations of the charge on the central electrode
and, in particular, the spectrum of the charge fluctuation as measured by
a detector coupled to the island (Section~\ref{long:sec3}), the counting
statistics for the tunneled charges (Section~\ref{long:sec4}), and the shot
noise of the current (Section~\ref{long:sec5}). In Section~\ref{long:sec5}
we discuss in some details the properties of the shot noise at zero frequency,
making explicit the results already contained in the counting statistics.
In the same section we also discuss the frequency dependence of the shot
noise. Section~\ref{long:sec6} is devoted to the conclusions.

\section{The model}
\label{long:sec2}

\begin{figure}
\includegraphics[width=.50\linewidth]{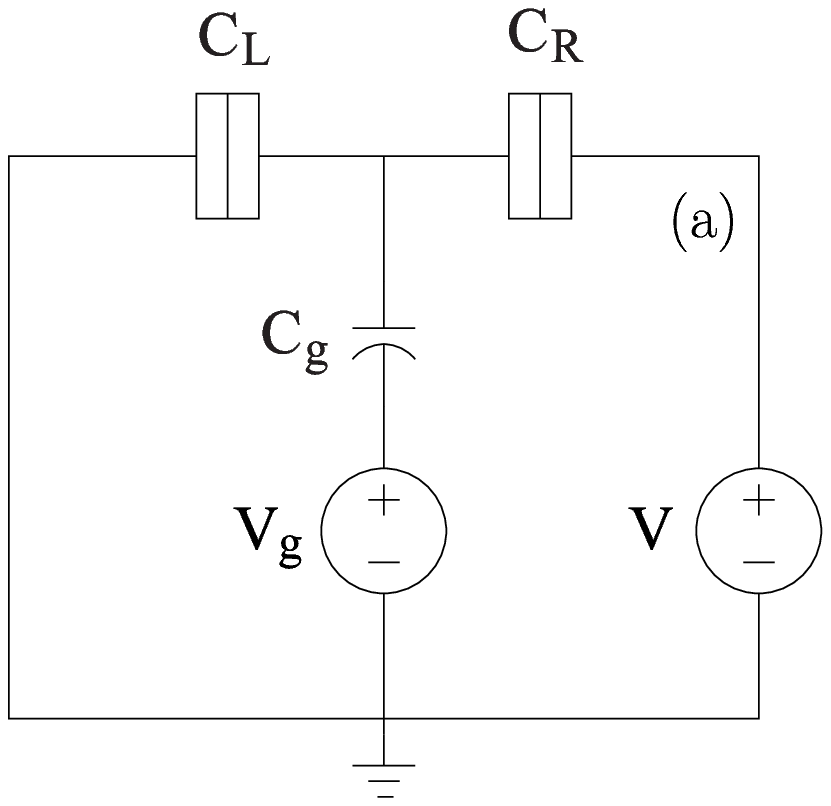}\hfill
\includegraphics[width=.45\linewidth]{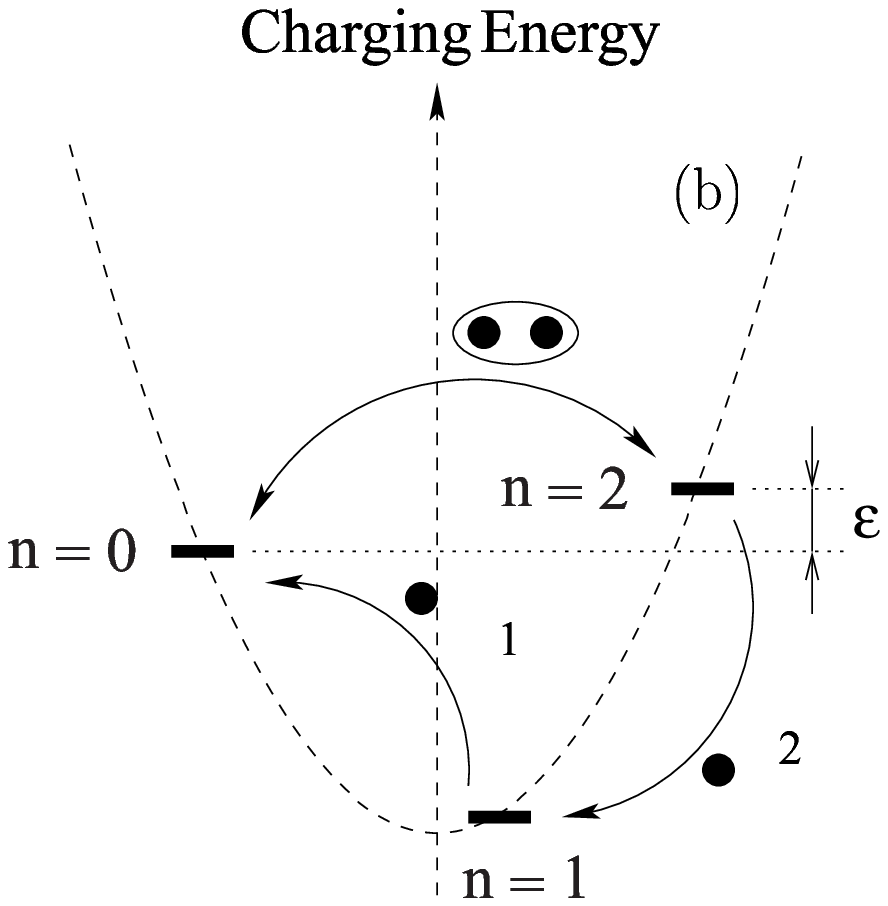}
\caption{Schematic
  diagrams of (a) the superconducting SET device and (b) the
  transition processes between relevant charge states.}
\label{fig01}
\end{figure}

The system we consider is a superconducting SET (see
Fig.~\ref{fig01}), which consists in a small central electrode
(island) connected by tunnel junctions to two leads and
capacitively coupled to a gate electrode. The electrostatic energy
of the island can be adjusted by controlling the gate voltage
$V_g$.  A transport voltage $V$ is applied to the outer leads,
determining the current flowing through the device.  Letting $C_g,
C_L$ and $C_R$ be the gate and left and right junction
capacitances, respectively, the electrostatic charging energy is
given by $E_C=e^2/2 C_{\Sigma}$ where $C_{\Sigma}=C_g+ C_L+C_R$ is
the total capacitance of the island.  The device operates in the
Coulomb blockade regime, i.e. with the charging energy much larger
than both the Josephson coupling energy $E_J$ and the thermal
energy $k_B T$ ($E_C\gg E_J, k_BT$), but still much smaller than
the superconducting gap $\Delta$ which we suppose to be the
largest energy scale in the problem. Charge can pass through the
tunnel barriers coherently and incoherently. In addition to
quasi-particle tunneling, Josephson effect allows to maintain
coherence in the Cooper-pair tunneling processes. This coherence
does play an important role in transport as long as charge states
which differ by one Cooper pair are almost degenerate. The
Hamiltonian of the system is given by
\begin{equation}
H_\mathrm{tot}=H_L+ H_R+ H_I+ H_T + H_C
\end{equation}
where $H_\alpha$ ($\alpha=L,R,I$) is the BCS Hamiltonian of the
left (L), right (R) lead  and of the central island (I). The
tunneling Hamiltonian~\cite{Ingold92} is
\begin{equation} \label{HT}
    H_T
    = \sum_{j=L,R}\sum_{kq\sigma}\left[ T_{kq} e^{-i\phi_j}\,
    c_{kj\sigma}^\dag
    c_{qI\sigma}^{} + h.c.\right] \,,
\end{equation}
where $T_{kq}$ is the tunneling amplitude  and $c_{k\alpha}^\dag$
($c_{k\alpha}$) creates (annihilates) a particle with momentum $k$
and energy $\varepsilon_{k\alpha}$ in electrode $\alpha$. The
variable $\phi_{L,R}$ is the superconducting phase difference at
the left (right) junction and it is canonically conjugated
 to the number $n_{L,R}$ of
electrons that have passed across the left (right)
junction~\emph{out of} the central electrode
($\left[n_k,\phi_j\right]=i\delta_{jk}$). Finally, $H_C$ is the
electrostatic energy, 
\begin{equation}
H_C=E_C (n+n_0)^2 + e V n_R \,, 
\end{equation}
where $n=-n_L-n_R$ is the number of excess electrons on the central
island, while $en_0\equiv{C_RV+C_gV_g}$ is the offset charge due
to the applied voltages.

For later convenience it is useful to define the part of the 
Hamiltonian which accounts for the coherent dynamics of the macroscopic 
variable $n$. It includes the charging and the Josephson terms. By
properly adjusting the bias and gate voltages, one can put either
the right or left junction at such a resonance for Cooper-pair
tunneling. We consider the case of resonance across the left
junction, and consequently we keep only the corresponding Josephson
coupling
\begin{equation}
\label{H0} H_0 = E_C(n+n_0)^2 + eVn_R - E_J\cos(2\phi_L) \,.
\end{equation}

A quasi-particle tunneling \emph{into} (\emph{out of}) the island across
the junctions leads to the transition $n\to{n+1}$ ($n\to{n-1}$) of the island
charge.  The rates of these incoherent processes are given by the
relation
\begin{equation} \label{Gamma}
\Gamma_{L/R}^\pm(n)
= \left[\coth(\beta\varE_{n,\pm}^{L/R})\pm 1\right]
  \frac{\im I_\mathrm{qp}(\varE_{n,\pm}^{L/R})}{2e} \,,
\end{equation}
where
\begin{math}
\varE_{n,\pm}^L=\pm{}E_{n,n\pm1}
\end{math},
\begin{math}
\varE_{n,\pm}^R=eV\pm{}E_{n,n\pm1}
\end{math},
\begin{math}
E_{m,n}=E_C(m-n)(m+n+2n_0)
\end{math},
and $I_\mathrm{qp}(E)$ is the quasi-particle tunneling current at the
bias voltage $E/e$ in the absence of charging
effects (see, e.g., Ref.~\onlinecite{Werthamer66} for an explicit
expression of $I_\mathrm{qp}$ in terms of $T_{kq}$).

Since the SET transistor operates in the charge regime ($E_C \gg
E_J$), we can take advantage of the strong suppression of charge
fluctuations to use the eigenstates of $n$ as basis states for the
island.  Moreover, we focus on the bias regime $|eV|\simeq
2\Delta+E_C$ where only the two charge states with $n=0$ and $n=2$,
are nearly degenerate. Such a condition implies that quasi-particle
tunneling only takes place from the central island toward the right
electrode, while the left junction allows only for coherent Cooper
pair tunneling.  Furthermore, we suppose that the Josephson energy of
the right junction is negligible (the corresponding term has already
been omitted from $H_0$, which is justified within the rotating wave
approximation). All these conditions are met in the recent experiment
by Nakamura \emph{et al.}~\cite{Nakamura99}, designed to probe the
state of the island via the detection of the incoherent tunneling
current.  In this situation one can imagine that the coherent
Cooper-pair tunneling occurring across the left junction is
interrupted from time to time by quasi-particle tunneling across the
right junction, as sketched in Fig.\ref{fig01} (b).

Due to the strong Coulomb blockade, it suffices to keep the three
charge states, $n=0,1,2$, and two tunneling rates,
$\Gamma_1\equiv\Gamma_R^-(1)$ and $\Gamma_2\equiv\Gamma_R^-(2)$; the
other tunneling rates are exponentially suppressed.  In order to simplify
the notation, we assume that $\Gamma_1=\Gamma_2\equiv\Gamma$, which is a
very good approximation in the regime we are interested in. For
example, in the experiment of Ref.~\onlinecite{Nakamura99},
$1/\Gamma_1 = 8 \mbox{ns}$ and $1/\Gamma_2=6 \mbox{ns}$.

The transport properties of the system in the set-up described above
can be well described in terms of two variables, either $n$ and $n_L$ or
$n$ and $n_R$ ($n=-n_L-n_R$).  However, the quantum dynamics of these system
is affected by the quantum noise due to the coupling to the environment
provided by the fermionic bath.  In order to describe
this effect, we adopt a master equation approach, which has been
widely used to describe quantum open systems~\cite{Carmichael93}. A
master equation for the reduced density matrix
$\rho(t)=\tr_\mathrm{qp}\rho_\mathrm{tot}(t)$ is obtained by taking
the trace over the fermionic degrees of freedom from the
Liouville equation ($\hbar=1$)
\begin{equation}
\label{long:Liouville}
\partial_t\rho_\mathrm{tot}(t)
= -i\left[H_\mathrm{tot},\rho_\mathrm{tot}(t)\right]
\end{equation}
for the density matrix $\rho_\mathrm{tot}$ of the system
plus environment.
The resulting equation can then be written in the Lindblad form
as~\cite{Averin89,vandenBrink91,Carmichael93,ChoiMS01a,ChoiMS01c}
\begin{widetext}
\begin{equation}
\label{long:MasterEq1}
\partial_t\rho(t)
 =  -i\left[H_0,\rho(t)\right]
  + \frac{1}{2}\sum_{n=1,2}\Gamma_n\Bigl[
      2L_n\,\rho(t)\,L_n^\dag
      - L_n^\dag L_n\,\rho(t)
      - \rho(t)\,L_n^\dag L_n
    \Bigr] \,.
\end{equation}
\end{widetext}
Here $L_n$ is a Lindblad operator corresponding to the quantum jump
$n\to{n-1}$ and $n_R\to{n_R+1}$, i.e., in the $\ket{n,n_R}$-basis
$L_n=\ket{n-1,n_R+1} \bra{n,n_R}$.  The first term describes a purely
phase-coherent dynamics, while the second one is responsible for both
dephasing and relaxation due to the quasi-particle tunneling.

The solution of the Eq.  \eqref{long:MasterEq1} behaves in distinct
ways in the two limiting cases of strong and weak coupling with the
quasi-particle reservoir.  In the \emph{strong dephasing limit}
(either $\Gamma \gg E_J$ or $\veps \gg E_J$, see below), the dephasing time
$\tau_\varphi$, which describes the decay of the off-diagonal elements
of $\rho$ to their stationary values, is small compared to the
relaxation time $\tau_r$ which sets the time-scale for the variation of
the diagonal elements (i.e. population of the charge states).
The relaxation time is given by
\begin{equation}
\frac{1}{\tau_r}
= \Gamma_\mathrm{r}
= \frac{2 E_J^2 \Gamma}{4 \veps^2 + \Gamma^2} \, .
\end{equation}
On the other hand, in the \emph{weak dephasing limit}
($\Gamma,\veps\ll{}E_J$), there is no such a clear separation of time
scales; both the diagonal and off-diagonal elements vary over the same time
scale $1/\Gamma$.

\section{Fluctuations of the charge on the island}
\label{long:sec3}

A first insight into the interplay among coherent Cooper-pair
tunneling, Coulomb blockade, and incoherent quasi-particle tunneling
can be obtained by examining the fluctuations of the charges on the
island as a function of the quasi-particle tunneling rate $\Gamma$,
the gate voltage, and the Josephson coupling energy.  In this case, we
only need to keep track of the variable $n$, and thus define a reduced
density matrix for the central island charge,
$\sigma(t)=\tr_{n_R}\rho(t)$, which satisfies an equation identical to
Eq. \eqref{long:MasterEq1}, but with $L_n=\ket{n-1}\bra{n}$ now
operating on the reduced `$n$' space only~\cite{ChoiMS01a}.  In the
stationary state, the master equation has the solution
($\sigma_{mn}=\bra{m}\sigma\ket{n}$)
\begin{subequations}
\label{long:sigma}
\begin{align}
\sigma_{00} & = \frac{1+ (4\veps^2+\Gamma^2)/E_J^2}
{3 + (4\veps^2+\Gamma^2)/E_J^2} \\
\sigma_{11} & = \sigma_{22} = \frac{1}
{3 + (4\veps^2+\Gamma^2)/E_J^2} \\
\sigma_{02} & = \sigma_{20}^* =
-i\frac{E_J(\sigma_{00}-\sigma_{22})}{\Gamma + 2i\veps}
\end{align}
\end{subequations}
Here $\veps=4 E_C (1+n_0)$ measures the energy difference between the
state with $n=0$ and $n=2$ charge on the island; the Cooper pair resonance
corresponds to $\veps=0$.

\begin{figure}
\includegraphics[width=.48\linewidth]{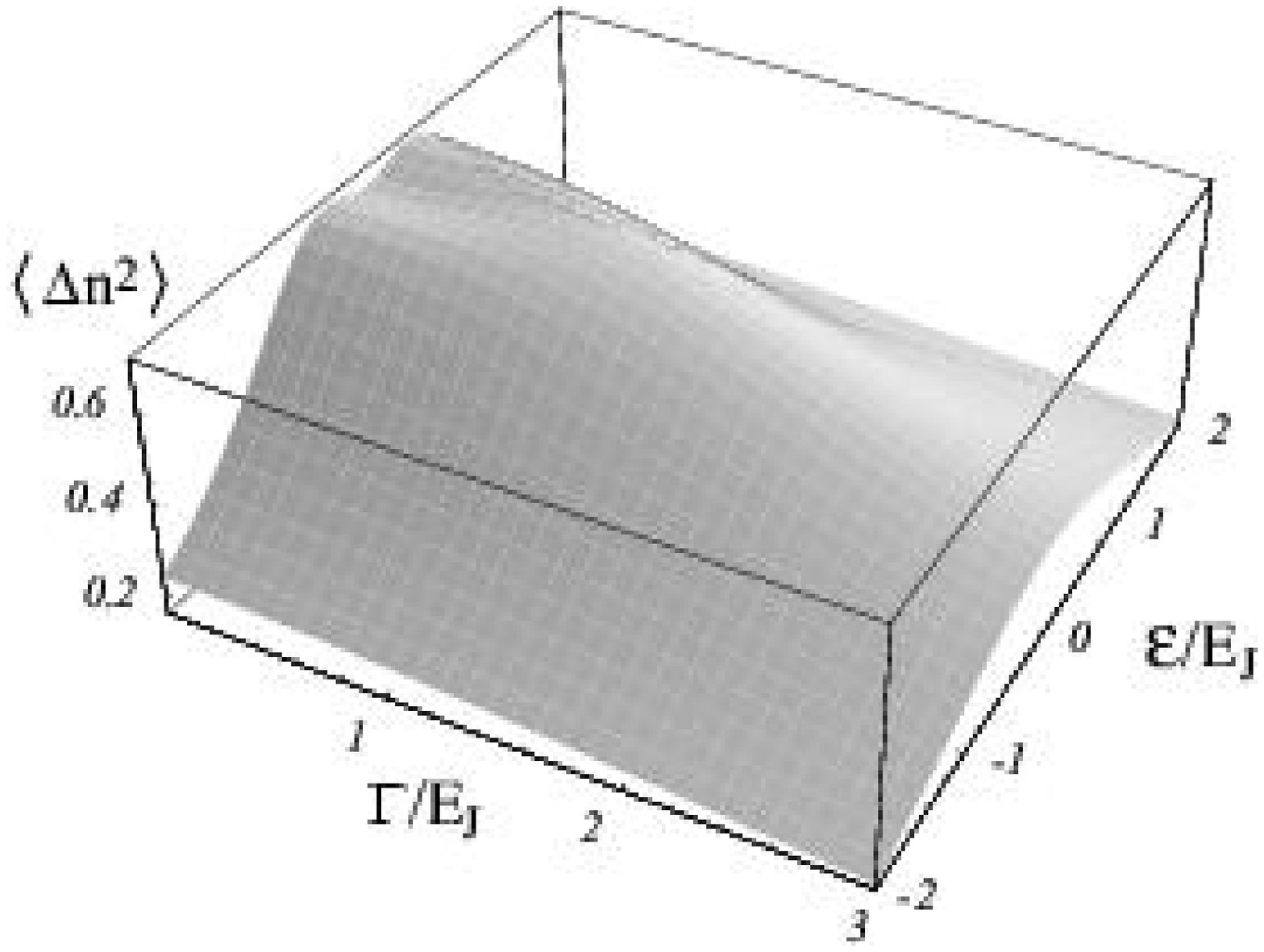}\quad\
\includegraphics[width=.35\linewidth]{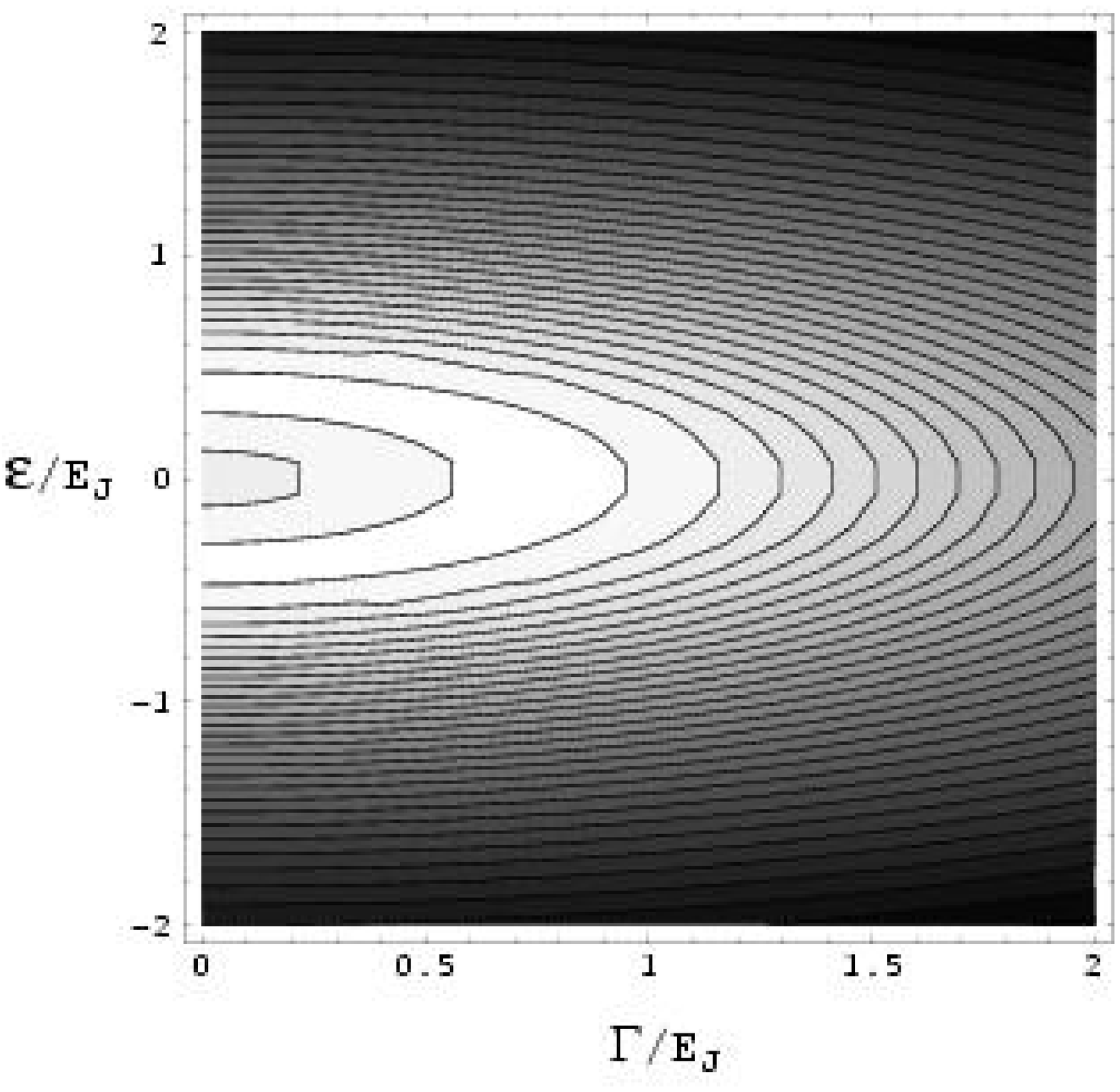}
\caption{Variance $\lavg(\Delta{n})^2\ravg$ of the charge on the island
  as a function of $\Gamma_1=\Gamma_2=\Gamma$ and $\veps$.}
\label{fig02}
\end{figure}

From the stationary-state solution \eqref{long:sigma}, we evaluate the
characteristic function
\begin{math}
C(\theta) = \left\langle e^{-i n \theta} \right\rangle
\end{math}
for the quantum variable $n$. It is given by
\begin{equation}
C(\theta)
= \frac{E_J^2 (e^{-2i \theta}+ e^{-i \theta}
   +1) + 4 \veps^2 + \Gamma^2}{4 \veps^2 +
    \Gamma^2 + 3 E_J^2 } \label{char}
\end{equation}
From $C(\theta)$ one can evaluate all the statistical moments of the
charge on the island; in particular, we concentrate on the variance
\begin{equation}
\label{long:variance}
\lavg(\Delta{n})^2\ravg
= E_J^2 \frac{ 5 (4 \veps^2 + \Gamma^2)
  + 6 E_J^2}
  {\left[4\veps^2 + \Gamma^2 + 3 E_J^2 \right]^2}
\end{equation}
shown in Fig.~\ref{fig02}.

An interesting point about the result in Eq.~\eqref{long:variance} is
that the maximum of the variance is found at
\begin{math}
\Gamma_\mathrm{opt} \simeq \sqrt{3/5}\,E_J
\end{math}
for $\veps=0$.  In other words, the fluctuation is enhanced when the
decoherence time matches the time for the Josephson coherent
oscillations.  Moreover as $|\veps|$ increases (up to
$|\veps|<E_J\sqrt{3/20}$), the optimal value of $\Gamma$ decreases as
\begin{math}
\Gamma_\mathrm{opt} \simeq \sqrt{3 E_J^2/5- 4 \veps^2}
\end{math}; meaning that
$\veps$ enhances the \emph{effective} dephasing rates.  These two
features will appear more clearly when we discuss the statistics
(Section~\ref{long:sec4}) and noise (Section~\ref{long:sec5}) of the
transport across the junctions.

An important quantity to consider is the fluctuation spectrum for the
number of electron charge residing on the island. As already discussed
by many authors, it is important to include the back-action of the
measuring apparatus~\cite{Makhlin00a,Devoret00,Averin00}, which
could be a SET transistor capacitively coupled to the central island.
We include the charge detector coupling via an Hamiltonian term
of the form
$$
\delta H_{det} = - \hat n \, D
$$
where $D$ is a detector operator. Assuming the correlation time
for the detector to be the fastest time scale of the problem, we write
(here we follow Averin's treatment~\cite{Averin00} ):
\begin{equation}
\lavg D(t+\tau) \, D(t) \ravg = \gamma_d \, \delta(\tau) \; .
\end{equation}
The non-zero value of $\gamma_d$ is the essential cause of the measurement
back-action. Indeed, a term proportional to $\gamma_d$
enters the master equation Eq.(\ref{long:MasterEq1}), thus affecting the time
evolution of the system variables. To effectively measure the charge number,
we look at an output detector operator $O$, which, in the linear regime,
evolves as $O(t) = O^{(0)}+ \lambda \hat n(t)$, (see Ref.\onlinecite{Averin00}).
The response coefficient $\lambda$ is determined by the imaginary part of  the equilibrium
correlation function $\lavg O(t+ \tau) D(t) \ravg$.
Furthermore, $\lambda$ can be related to
$\gamma_d$ so that we can write for the signal to noise ratio
\begin{equation}
\frac{S_O (\omega)}{S_{noise}}= 1+ 2 \gamma_d \, S_n(\omega)
\end{equation}
where it is assumed that the real part of the $O$--$D$ correlator  vanishes
(which is the most favorable case for a measurement).
Here $S_n(\omega)$ is the charge number fluctuation spectrum evaluated as
\begin{equation}
S_n(\omega) = \frac{1}{2} \lim_{t \rightarrow \infty}
\int_{-\infty}^{+\infty} \lavg \Bigl \{ \hat n(t+\tau) ,
\hat n(t) \Bigr \}\ravg \, e^{i \omega \tau} \, d \tau
\end{equation}
where the time evolution is obtained from a modified master equation
including the back-action. Note that here we use the symmetric correlation
function since the island charge itself is coupled
to the quasi-particle bath, so that the detector can also receive energy
from the system.

\begin{figure}
\includegraphics[width=.42\linewidth]{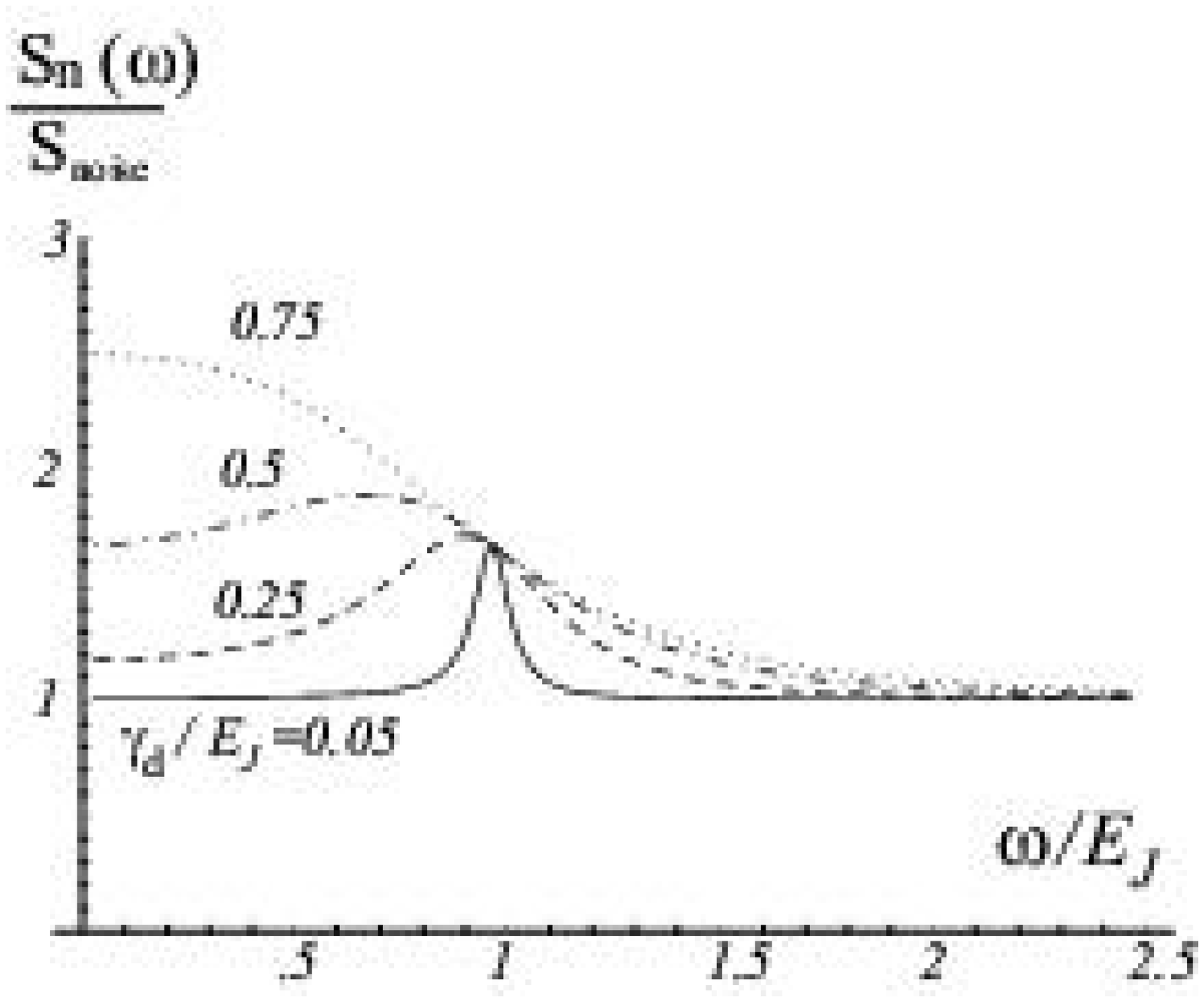}\quad\
\includegraphics[width=.42\linewidth]{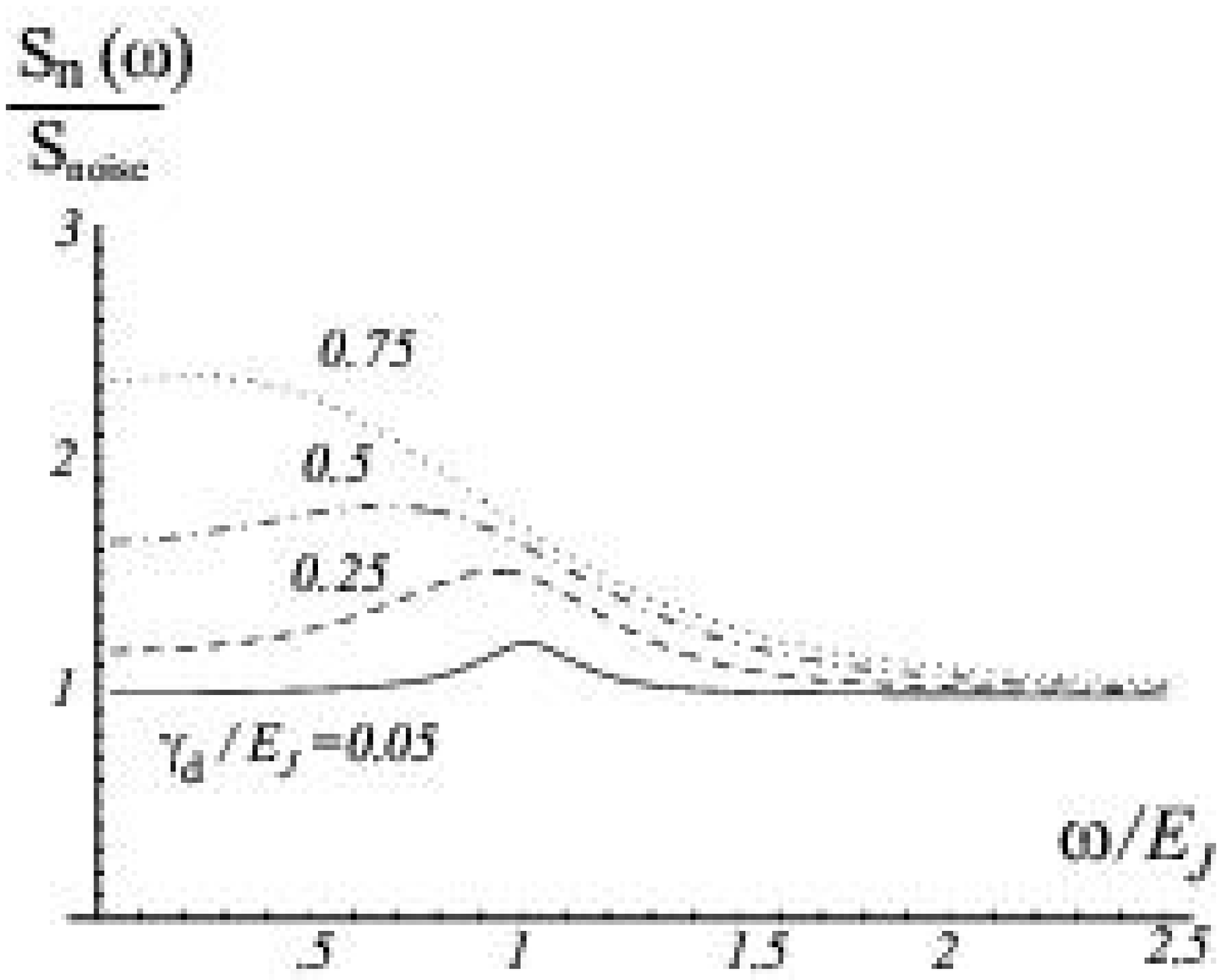}
\caption{Charge fluctuation spectrum in the weak dephasing regime at degeneracy ($\veps=0$),
with the indicated back-action rate values and with $\Gamma/E_J=10^{-5}$ (left) and  $\Gamma/ E_J=0.1$ (right).}
\label{fig03}
\end{figure}

As shown in Fig. (\ref{fig03}), the spectrum displays a resonance peak
at the Josephson frequency which is broadened by the quasi-particle rate
$\Gamma$ (the peak is only visible in the weak dephasing regime, otherwise
it is completely washed out independently of the value of $\gamma_d$).
As the back-action rate $\gamma_d$ increases, a maximum develops at zero
frequency, which finally hides the resonance structure. This enhanced zero
frequency  noise results from incoherent transition induced by the detector
coupling.

Near the maximum at $\omega\simeq E_J$, and for $\Gamma, \gamma_d \ll E_J$,
the spectrum takes the approximate form
\begin{equation}
S_n(\omega)  \simeq \frac{8}{3} E_j^2 \, \frac{(\Gamma+ 2 \gamma_d)}
{4 E_J^2(\Gamma + 2 \gamma_d)^2 +
(\omega^2- E_j^2)^2} \; .
\end{equation}

\section{Counting statistics for the transmitted charge}
\label{long:sec4}

\begin{figure}
\includegraphics[width=.48\linewidth]{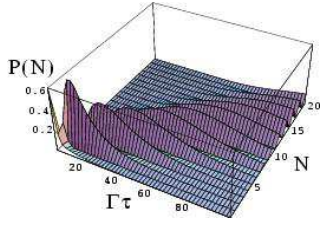}
\caption{$P_{t\approx0}(N, \tau)$ obtained by
  numerically solving equation (\ref{long:MasterEq2}), in the case of
  $\Gamma_\mathrm{r}=0.1 \Gamma$, with initial condition
  $\rho_{00}(0)=1, \rho_{22}(0)=0$.}
\label{fig04}
\end{figure}

\begin{figure}
\includegraphics[width=.48\linewidth]{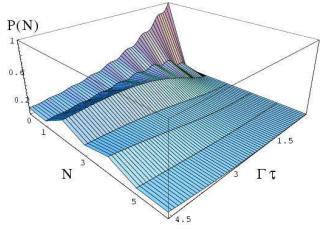}
\caption{The short time probability for the first few tunneling events
  in the weak dephasing regime ($\Gamma=0.1 E_j$) and at resonance
  ($\veps=0$). In this regime, the counting probability does not depend on the initial condition for $\rho$}
\label{fig05}
\end{figure}

We turn now to a description of the statistical distribution of
the number of charges transmitted through the system during a
period $\tau$, \cite{levitovcount}.  Specifically, we will examine
the probability $P_t(N,\tau)$ that $N$ electrons have been
transferred across the right junction during the interval
$[t,t+\tau]$.  We note that
\begin{equation}
\label{long:P-p}
P_t(N,\tau) = \sum_{n_R} p(N+n_R, t+\tau;n_R,t) \,,
\end{equation}
where $p(n_R+N,t+\tau;n_R,t)$ is the joint probability that $n_R$
electrons have passed across the right junction up to the time $t$ and
$n_R+N$ electrons up to time $t+\tau$.  To obtain $P_t(N,\tau)$, we
define a characteristic matrix
\begin{widetext}
\begin{equation}
\label{long:G}
G_t(\theta,\tau)
= \sum_{n_R,N}\exp\left(-i\theta N\right)\tr_\mathrm{qp}
  \bra{n_R+N}e^{-iH_\mathrm{tot}\tau}\ket{n_R}\bra{n_R}
  \rho_\mathrm{tot}e^{+iH_\mathrm{tot}\tau}\ket{n_R+N} \, ,
\end{equation}
defined so that $\tr{}G_t(\theta,\tau)\equiv \lavg e^{i \theta N}
\ravg$ is the characteristic function for $P_t(N,\tau)$. Namely,
\begin{equation}
P_t(N,\tau)
= \int_{-\pi}^{\pi}\frac{d\theta}{2\pi}\;
  e^{+i\theta N} \tr G_t(\theta,\tau) \,,
\label{serve}
\end{equation}
where the trace is taken over the states $\ket{n}$.  Following the
same procedure that led to Eq.~\eqref{long:MasterEq1}, one can show
that $G_t(\theta,\tau)$ satisfies the following master equation:
\begin{equation}
\label{long:MasterEq2}
\partial_{\tau} G_t
= -i\left[H_0,G_t\right]
  + \frac{1}{2}\sum_{n=1,2}\Gamma_n
    \left[ 2 e^{i \theta} L_n G_t L_n^{\dag}
    - G_t L_n^{\dag} L_n
    - L_n^{\dag} L_n G_t
    \right]
\end{equation}
with the initial condition
\begin{math}
G_t(\theta,0) = \sum_{n_R}\bra{n_R}\rho(t)\ket{n_R}
\end{math}.

Here we will consider two limiting cases for the solution, the strong
and the weak dephasing limit (see the discussion at the end of
Section~\ref{long:sec2}).  We find that in the strong dephasing case
($\Gamma\gg{}E_J$ or $\veps\gg{}E_J$)
\begin{equation}
\label{long:G1}
\tr G_t(\theta,\tau)
\simeq \left[\sigma_{00}(t) + z\sigma_{11}(t) + z^2 \sigma_{22}(t)
  \right]
  \exp\left[- \frac{\Gamma_\mathrm{r}\tau}{2} (1-z^2)\right]
\end{equation}
while in the weak dephasing limit at resonance ($\Gamma \ll E_J, \, \veps=0 $)
\begin{multline}
\label{long:G2}
\tr G_t(\theta,\tau) = \Biggl\{ \left[\cosh\frac{\Gamma\tau f(z)}{4} +
\frac{1+2\sigma_{11}(t) + 2z[1-\sigma_{11}(t)]}{f(z)}
\sinh\frac{\Gamma\tau f(z)}{4}
\right] \\
+ \frac{\Gamma}{E_J}(1-z)\im\sigma_{02}(t) \left[\cosh\frac{\Gamma\tau
  f(z)}{4} + \frac{4z+1}{f(z)}\sinh\frac{\Gamma\tau f(z)}{4} \right]
\Biggr\} \exp\left(-\frac{3\Gamma\tau}{4}\right) \\
- \frac{\Gamma}{E_J}(1-z)\left\{ \im\sigma_{02}(t)\cos(E_J\tau) -
\frac{1}{2}\left[\sigma_{00}(t)-\sigma_{22}(t)\right]\sin(E_J\tau)
\right\} \exp\left(-\frac{\Gamma\tau}{2}\right) \,,
\end{multline}
\end{widetext}
where $z=e^{i\theta}$, and $f(z)=\sqrt{1+8z^2}$.  In Figs.~\ref{fig04}
and \ref{fig05}, the resulting statistics are shown for the weak and
strong dephasing cases, respectively, in the transient state (i.e.
$t\simeq 0$).

\begin{figure}
\includegraphics[width=6cm]{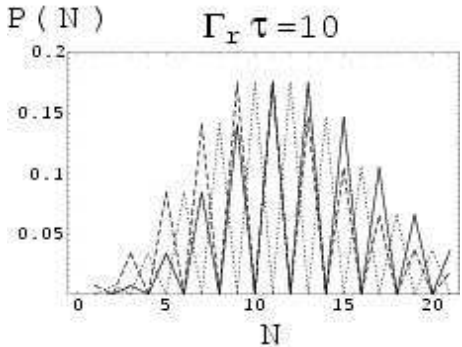}
\caption{Comparison among transient-state counting statistics
  $P_{t=0}(N,\tau)$ at $\Gamma_r\tau=10$ in the strong dephasing
  regime with $\Gamma_r/\Gamma = 0.1$ for the three initial conditions
  a) $\sigma_{00}=1, \sigma_{11}=\sigma_{22}=0$ (dashed line), b)
  $\sigma_{11}=1, \sigma_{00}=\sigma_{22}=0$ (dotted line) c)
  $\sigma_{22}=1, \sigma_{00}=\sigma_{11}=0$ (continuous line) .}
\label{fig06}
\end{figure}

As seen in Eqs.~\eqref{long:G1} and \eqref{long:G2},
$G_t(\theta,\tau)$ and hence $P_t(N,\tau)$ depend on the charge
state of the island at time $t$.  This point is further illustrated in
Fig.~\ref{fig06} where the counting statistics at short time is shown
in the strong dephasing limit for various starting conditions
$\sigma(t)$.
%We would like to emphasize that this is not a phenomenon
%for merely a theoretical consideration since recent advances of
%quantum technology allow to coherently control the quantum states of
%such devices and to experimentally prepare an arbitrary state
%$\sigma(t)$.
In the limit of strong dephasing, the counting
statistics depends sensitively on the initial state $\sigma(t)$. This
has been exploited in Ref.~\onlinecite{Nakamura99}, where the
measurement of the quantum state is performed with the system taken
far from degeneracy.  On the contrary, the dependence on the initial
condition is quickly lost in the weak dephasing regime, since the
strong Josephson energy can rapidly produce a change in the state,
before quasi-particles have any time to be produced.

Another important limit to consider is the stationary state
($t\to\infty$), where physical properties do not depend on the initial
preparation of the system.  In the strong dephasing limit,
Eq.~\eqref{long:G1} is reduced to the simple form
\begin{equation}
\tr G_\infty(\theta,\tau)
\simeq \exp\left[-\half\Gamma_\mathrm{r}\tau(1-z^2)\right] \,.
\end{equation}
It gives the probability distribution function for the transmitted
charges
\begin{subequations}
\label{countstat}
\begin{align}
\label{countstata}
& P_\infty(2 N + 1, \tau) = 0 \\
\label{countstatb}
& P_\infty(2 N,\tau) = \frac{(\Gamma_\mathrm{r} \tau /2)^{N}}{N !}
\; \exp \left(- \frac{\Gamma_\mathrm{r}\tau}{2}\right) \,.
\end{align}
\end{subequations}
$P_{\infty}(N)$ shows a strong even-odd asymmetry: for even $N$, the
distribution is Poissonian, but the probability that an odd number of
electrons has passed is negligible.  Below we will see that this
strong parity effect manifests itself as an enhancement of
zero-frequency shot noise.  We leave the physical interpretation of
the parity effect until we discuss shot noise in
Section~\ref{long:sec5}.

In the weak dephasing limit, Eq.~\eqref{long:G2} is reduced to
%\begin{widetext}
\begin{multline}
\tr G_\infty(\theta,\tau) = \exp\left(-\frac{3 \Gamma \tau}{4}\right)
\\ \mbox{}
\times \left[ \cosh \frac{\Gamma \tau f(z)}{4} + \frac{1+ 8 z}{3
  f(z)} \sinh \frac{\Gamma \tau f(z)}{4} \right]
\end{multline}
%\end{widetext}
so that
\begin{subequations}
\label{P1}
\begin{align} \label{P1a}
& P_\infty(2N,\tau) = \exp\left(-\frac{3\Gamma\tau}{4}\right)
\left(\frac{1}{3}+\frac{4}{\Gamma}\frac{\partial}{\partial\tau}\right)
F_{N}(\tau) \,, \\
& P_\infty(2N-1,\tau) =
\frac{8}{3}\exp\left(-\frac{3\Gamma\tau}{4}\right)\, F_{N}(\tau) \,,
\label{noise:P1b}
\end{align}
\end{subequations}
where
\begin{equation}
F_n(\tau) = \frac{1}{2\pi i}\oint_{|z|=1}\frac{dz}{z^{n+1}}\;
  \frac{1}{f(z)}\sinh\frac{\Gamma\tau f(z)}{4} \,.
\end{equation}
This distribution function shows a much weaker (but still finite)
even-odd asymmetry than the previous case [cf. Eq.~(\ref{countstat})].
Furthermore, the distribution clearly deviates from a Poissonian
function, indicating that the presence of the strong coherent
tunneling of Cooper pairs tends to correlate the quasi-particle
tunneling events across the right junction.  This is further reflected
in the deviations of the current noise from the classical shot noise
value (see discussions in Section~\ref{long:sec5}).

\begin{figure}
\includegraphics[width=6cm]{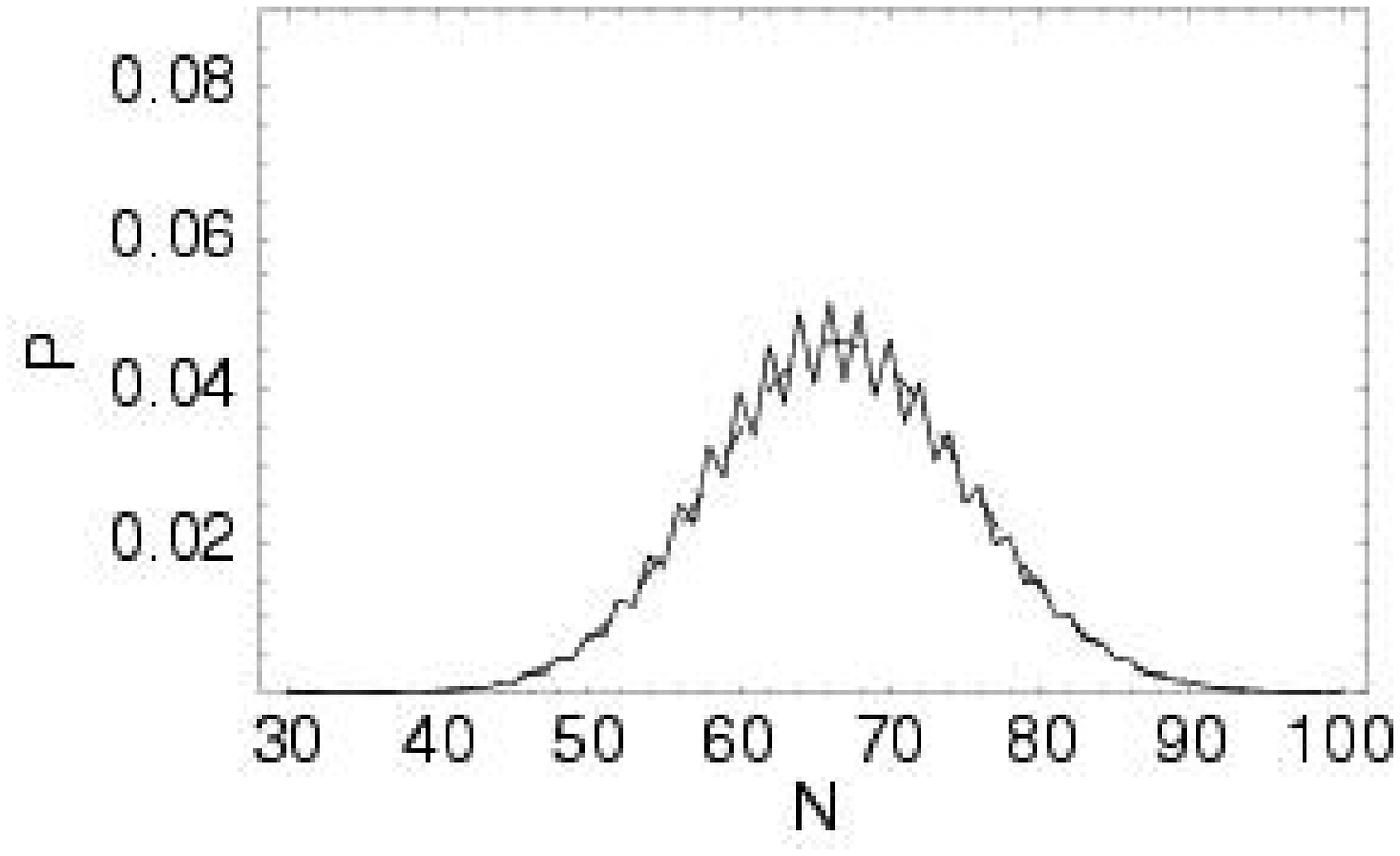}
\includegraphics[width=6cm]{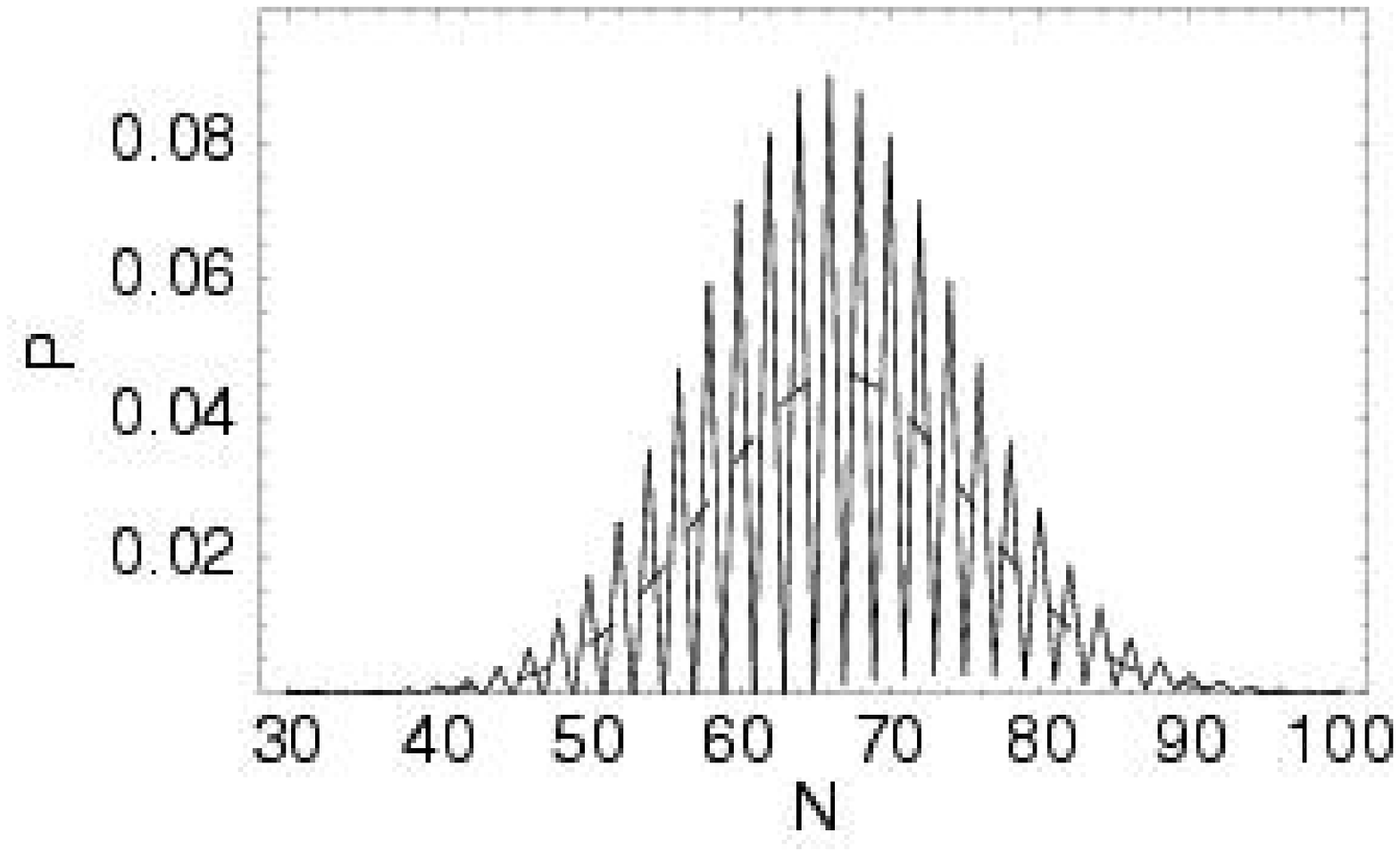}
\caption{Stationary-state distribution $P_\infty(N,\tau)$ at
  $\Gamma\tau=100$ (a) $\Gamma_1=\Gamma_2=\Gamma=10^{-5}E_J$ and (b)
  $\Gamma_1=\Gamma_2=\Gamma=30E_J$.  For a comparison, a normal
  distribution function given in Eq.~(\ref{normal}) is also shown
  (dashed line).}
\label{fig07}
\end{figure}

One may expect that for a long waiting time ($\Gamma\tau\to\infty$),
implying very large numbers of tunneled charges, the distribution of
$N$ should approach a Gaussian.  In particular, this becomes an exact
result if the distribution is Poissonian.  In our case, on the other
hand, we have
\begin{subequations}
\label{long:P2}
\begin{align}
& P_\infty (2N, \tau\to\infty) \approx \frac{5}{9}P_G(N,\tau) \,,\\
& P_\infty(2N-1, \tau\to\infty) \approx \frac{4}{9}P_G(N,\tau) \,,
\end{align}
\end{subequations}
where $P_G$ is a Gaussian distribution,
\begin{equation}
P_G(N, \tau)= \frac{1}{\sqrt{2 \pi \eta \tau}}
\exp \left [-\frac{(N- I \tau/2e)^2}{2 \eta \tau} \right ]
\label{normal}
\end{equation}
with $\eta= 20/27$.  The distributions for both even and odd $N$ are
separately Gaussian, but $P_\infty(N, \tau \gg \Gamma^{-1})$ as a
whole is not, since even-odd asymmetry is still present.  In
Fig.~\ref{fig07} we compare the stationary-state  results for
$P_\infty(N,\tau)$ in the weak and strong dephasing limits with the
Gaussian distribution.

\begin{figure}
\includegraphics[width=.59\linewidth]{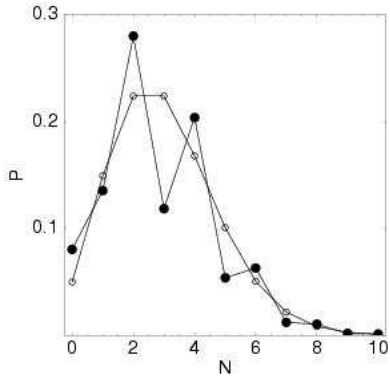}
\caption{Counting probability $P_\infty(N,\tau)$ at $\Gamma\tau=4$ for
  $\Gamma=\sqrt{2}E_J$ (filled circle). For a comparison, the
  Poissonian distribution is also plotted (empty circle).}
\label{fig08}
\end{figure}

Finally, it is interesting to understand what happens to the
stationary counting probability $P_\infty(N, \tau)$ in the
intermediate regime, i.e., when the dephasing rate is comparable to
the Josephson energy. Unfortunately, an analytic expression is not
available in this case; the numerical results, however, are shown in
Fig.\ref{fig08}, where one can see that the distribution function
deviates significantly from a Poissonian, being suppressed (enhanced)
for odd (even) $N$.

\section{Shot Noise}
\label{long:sec5}

A deeper insight into the transport process can be obtained in the
frequency domain, from a careful analysis of the spectral power of
current fluctuations. The zero-frequency shot noise could be directly
determined by the second moment of the counting probability
Eq.~(\ref{serve}), see Ref.~\onlinecite{Blanter00}. Here, however, we
follow a different route which allows us to get the entire current
spectrum.  To this end, we define the noise spectrum as
\begin{equation}
\label{Sdef}
S(\omega)
= \lim_{t \rightarrow \infty} \int_{-\infty}^\infty{d\tau}\;
  e^{i\omega\tau}
  \lavg\left\{\delta I(t+\tau),\delta I(t)\right\}\ravg \,,
\end{equation}
where $\delta{I}(t)=I(t)-\langle{I(t)}\rangle$ and $\{A,B\}=AB+BA$.
The total current $I(t)$ through the system is related to the
\emph{tunneling} currents $I_{L/R}=-e\partial_tn_{L/R}$ across each
junction by \cite{Korotkov94a}
\begin{equation}
\label{totalcurrent}
I(t) = \frac{C_L}{C_\Sigma}I_R(t) - \frac{C_R}{C_\Sigma}I_L(t) \;.
\end{equation}
To simplify the evaluation of $S(\omega)$, it is convenient to
introduce the spectral densities of currents flowing across the
individual junctions and the cross correlation spectral powers.
Therefore, in a way analogous to Eq. (\ref{Sdef}), we write
($i,j=L,R$)
\begin{equation} \label{Sijdef}
S_{ij}(\omega) = \lim_{t \rightarrow \infty}
\int_{-\infty}^\infty{d\tau}\;e^{i\omega\tau}
  \lavg\left\{\delta I_i(t+\tau),\delta I_j(t)\right\}\ravg ,
\end{equation}
which allows to express the total shot noise spectrum in the form
%\begin{widetext}
\begin{multline}
\label{Sdecomposition}
S(\omega)
 =  \frac{C_R^2}{C_\Sigma^2}S_{LL}(\omega)
  + \frac{C_L^2}{C_\Sigma^2}S_{RR}(\omega) \\ \mbox{}
 - \frac{C_LC_R}{C_\Sigma^2}\left[S_{LR}(\omega)+S_{RL}(\omega)\right] \,.
\end{multline}
%\end{widetext}
In the stationary state
$\lavg{I}\ravg=\lavg{I_L}\ravg=-\lavg{I_R}\ravg$, so that
$S(\omega)=S_{LL}(\omega)=S_{RR}(\omega)$ in the zero-frequency limit.
In the opposite limit ($\omega\to\infty$),
\begin{math}
S(\omega) = (C_L^2/C_\Sigma^2)S_{RR}(\omega) =
(C_L^2/C_\Sigma^2)\,2e\lavg{I}\ravg
\end{math} \cite{Davies92,Hershfield93,Korotkov94a}.
In our case, the left junction is (nearly) at resonance for the Cooper
pair tunneling and therefore $\lim_{\omega\to\infty}S_{LL}(\omega)=0$.

In order to obtain $S_{ij}(\omega)$ we have to calculate two-time
correlation functions. We follow the standard procedure based on the
quantum regression theorem~\cite{Carmichael93} and define the auxiliary
matrices
\begin{subequations}
\label{long:ChiEta}
\begin{align}
\label{long:Chi}
&\chi^{(j)}(t,\tau) = \tr_\mathrm{qp} \left\{ e^{-i H \tau}
n_j \rho_\mathrm{tot}(t) e^{i H\tau} \right\} \,,\\
\label{long:Eta}
&\eta^{(j)}(t,\tau) = \tr_\mathrm{qp} \left\{ e^{-i H \tau}
\rho_\mathrm{tot}(t) n_j e^{i H\tau} \right\} \,,
\end{align}
\end{subequations}
where the index $j$ runs over the left and right junctions ($j=L,R$).
These auxiliary operators, $\chi^{(j)}$ and $\eta^{(j)}$, satisfy a
master equation of exactly the same form as Eq.~\eqref{long:MasterEq1}
(but with respect to $\tau$ instead of $t$ and, of course, with
different initial conditions).  Their relevance can be understood by
noticing that the correlation functions can be expressed directly in
terms of their average values\cite{foot}:
\begin{widetext}
\begin{eqnarray}
\label{long:C-RR}
\lavg \left\{ \delta I_R(t+ \tau), \delta I_R(t) \right\} \ravg
= e^2 (\partial_{\tau}-\partial_t) \sum_{n_R}
  \sum_{n=1,2} &\Gamma_n &\bra{n,n_R}\chi^{(R)}(t,\tau)
    + \eta^{(R)}(t,\tau)\ket{n,n_R} \nonumber \\
 &-& 2\lavg I_R(t) \ravg^2
  - 2 e \lavg I_R(t) \ravg \delta(\tau) \,,
\end{eqnarray}
\begin{equation}
\label{long:C-LL}
\lavg \left\{ \delta I_L(t+ \tau), \delta I_L(t) \right \}\ravg
= 2 e^2 (\partial_t-\partial_{\tau})(\partial_{\tau}+ \Gamma)
  \sum_{n_R}\bra{2,n_R}\chi^{(L)}(t,\tau) +\eta^{(L)} (t,\tau)\ket{2,n_R}
  - 2 \lavg I_L(t)\ravg^2
%  + 2 e \lavg I_L(t) \ravg \delta(\tau) \,,
\end{equation}
and
\begin{multline}
\label{long:C-LR}
\lavg \left\{ \delta I_L(t+ \tau), \delta I_R(t) \right \}\ravg +
\lavg \left\{ \delta I_R(t+ \tau), \delta I_L(t) \right\} \ravg
= e^2 (\partial_t-\partial_{\tau})\sum_{n_R} \biggl \{ 2
(\partial_{\tau}+ \Gamma) \bra{2,n_R} \chi^{(R)} + \eta^{(R)}\\
\ket{2,n_R} - \sum_{n=1,2}\Gamma \bra{n,n_R} \chi^{(L)} + \eta^{(L)}
\ket{n,n_R} \biggr \}
- 4 \lavg I_L(t) \ravg \lavg I_R(t) \ravg
%+ 2e \left( \lavg I_L(t) \ravg + \lavg I_R(t) \ravg \right) \delta(\tau)
\,.
\end{multline}
The problem is now reduced to (a) solving a master equations of the
form given in Eq.~\eqref{long:MasterEq1} to get $\rho(t)$,
$\chi^{(j)}(t,\tau)$, and $\eta^{(j)}(t,\tau)$ for respective initial
conditions, and (b) evaluating Eqs.~\eqref{long:C-RR},
\eqref{long:C-LL}, and \eqref{long:C-LR} to obtain the correlation
functions.  Following this procedure and performing the Fourier
transforms of the resulting correlation functions, we find that in the
stationary state
\begin{equation}
\label{long:S-RR}
\frac{S_{RR}(\omega)}{2e \lavg I \ravg }
= 1 - \bra{A}\frac{\Gamma(M-2 \Gamma)}{\omega^2+ M^2}\ket{A} \,,
\end{equation}
\begin{equation}
\label{long:S-LL}
\frac{S_{LL}(\omega)}{2 e \lavg I \ravg}
= 2\bra{B}\frac{M-\Gamma}{\omega^2+M^2}
  (M\ket{C}-2\Gamma\ket{A}) \,,
\end{equation}
and
\begin{equation}
\label{long:S-LR}
\frac{ S_{LR}(\omega) + S_{RL}(\omega) }{2 e \lavg I \ravg}
= 2\bra{B}\frac{(M-\Gamma)(M-2\Gamma)}{\omega^2+M^2}\ket{A}
+ \bra{A}\frac{\Gamma}{\omega^2+M^2} (M\ket{C}-2\Gamma\ket{A}) \,,
\end{equation}
\end{widetext}
where we have used the bra-ket notations
\begin{equation}
M
= \begin{pmatrix}
  \Gamma & \Gamma & i E_J/2 & -i E_J/2
  \cr 0 & \Gamma & -i E_J/2 & i E_J/2 \cr i E_J/2 & -i E_J/2 &
  \Gamma/2 + i \veps & 0 \cr -i E_J/2 & i E_J/2 & 0 & \Gamma/2-i
  \veps
  \end{pmatrix} \,,
\end{equation}
\begin{equation}
\label{braketdef}
\ket{A}
= \begin{pmatrix}
  1 \cr 0 \cr
  0 \cr 0
  \end{pmatrix} \,,
\ket{B}
= \begin{pmatrix} 0 \cr 1 \cr 0 \cr 0\end{pmatrix} \,,
\end{equation}
and
\begin{equation}
\ket{C}
= \ket{A} + \ket{B}
+ \left(\frac{E_J}{\Gamma}+\frac{E_J}{\Gamma_\mathrm{r}}\right)
  \begin{pmatrix}
  0\\ 0\\ -i\\ +i
  \end{pmatrix} \,.
\end{equation}

\subsection{Zero-Frequency Noise}
\label{long:sec5.1}

%Noise spectrum at zero-frequency is particularly interesting since, usually,
%low-frequency parts of the noise spectrum are more easily
%obtained experimentally than high-frequency ones.  Moreover, the
%zero-frequency noise gives a direct information about the fluctuations
%in transported charges~\cite{Davies92,Hershfield93,Hanke93-94}.
From Eqs.~\eqref{long:S-RR}, \eqref{long:S-LL}, \eqref{long:S-LR}, and
\eqref{Sdecomposition}, it follows that the zero-frequency noise is
given by
\begin{equation}
\label{Szero}
\frac{S(0)}{2e \lavg I\ravg}
= 2 - \frac{8E_J^2(E_J^2+2\Gamma^2)}
       {(3E_J^2 + \Gamma^2 + 4\veps^2)^2} \;.
\end{equation}

In the strong dephasing limit ($\Gamma\gg{E_J}$), the second term in
Eq.~\eqref{Szero} becomes negligibly small, as it vanishes as
$(E_J/\Gamma)^2$.  Therefore, the zero-frequency shot noise is
enhanced approximately by a factor $2$ compared with its classical
value, $2e\lavg{I}\ravg$. This can be understood in terms of the
Josephson quasi-particle (JQP) cycle~\cite{Fulton89a,Averin89,vandenBrink91}.
Because of the fast quasi-particle tunneling across the right
junction, each Cooper pair that has tunneled into the central island
breaks up immediately into quasi-particles, and quickly tunnels out.
The charge is therefore transferred in units of $2e$ (compared with
$e$ in classical charge transfer) for each JQP cycle.  This was
already confirmed in the counting statistics of the transmitted
charges.  According to Eq.~\eqref{countstat}, the probability that an
odd number of electrons are transferred is zero and charges are
transferred only in pairs. In the weak ($\Gamma\ll{E_J}$) and moderate
($\Gamma\simeq{E_J}$) dephasing limits, the semiclassical JQP picture
breaks down and we do not have shot noise enhancement any longer.

In the limit $\Gamma\ll{E_J}$, the period of oscillations of Cooper pair
is very short compared to the typical time for for quasi-particles to
tunnel out of the central island.  The
system can be regarded as a single-junction circuit, where the
quasi-particle tunneling events are independent.  The Fano factor
$S(0)/2e\lavg{I}\ravg\approx10/9$,  becomes much closer to
unity in this limit.  The small deviation from the Poisson value
is due to the fact that  the quasi-particle tunneling events cannot
be considered as independent because Coulomb blockade allows only one
Cooper pair to oscillate coherently across the left junction.
Therefore, the tunneling process corresponding to $\ket{2}\to\ket{1}$
is likely to be followed by $\ket{1}\to\ket{0}$. It is clear that this
behavior is related to the residual even-odd asymmetry we found in the
counting statistics, Eq.~\eqref{P1}, even in the long waiting-time limit
($\Gamma\tau\gg1$), Eq.~\eqref{long:P2}.

\begin{figure}
\includegraphics[width=6cm]{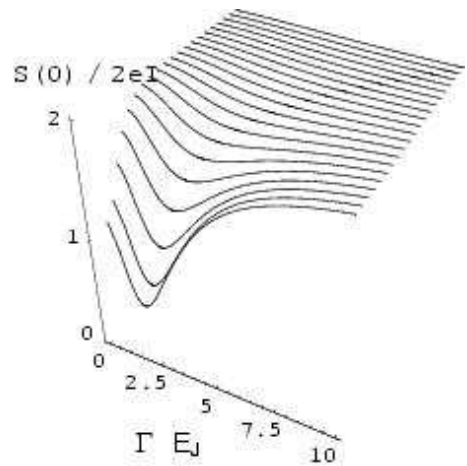}
\caption{Normalized zero-frequency shot noise for
  $\veps/E_J=0,0.25,\ldots,5$.  The dip in the noise is most
  pronounced at resonance ($\veps=0$).}
\label{fig09}
\end{figure}

With moderate dephasing ($\Gamma\simeq{E_J}$), quasi-particle
tunneling events across the right junction are strongly affected by
the \emph{coherent\/} oscillation of Cooper pairs across the left
junction. Indeed, this effect gives rise to the significant deviation
from the Poissonian distribution of the tunneling statistics,
Eq.~\eqref{P1}.  Most remarkably, it leads to a suppression of the
shot noise.  The strongest suppression, by a factor of $2/5$, is
achieved at resonance ($\veps=0$) for $\Gamma=\sqrt{2}E_J$, see
Fig.~\ref{fig09}.  This is reminiscent of the shot noise suppression
in (non-superconducting) double-junction systems\cite{Hershfield93},
whose maximal suppression is by factor $1/2$ for the symmetric
junctions.  We emphasize, however, that in the latter case, the
coherence was not essential. In our case, on the contrary, the role of
coherence becomes evident by noticing that the dip in Fano factor
disappears when moving away from the resonant condition as shown in
Fig.~\ref{fig09}.

\begin{figure}
\includegraphics[width=.45\linewidth]{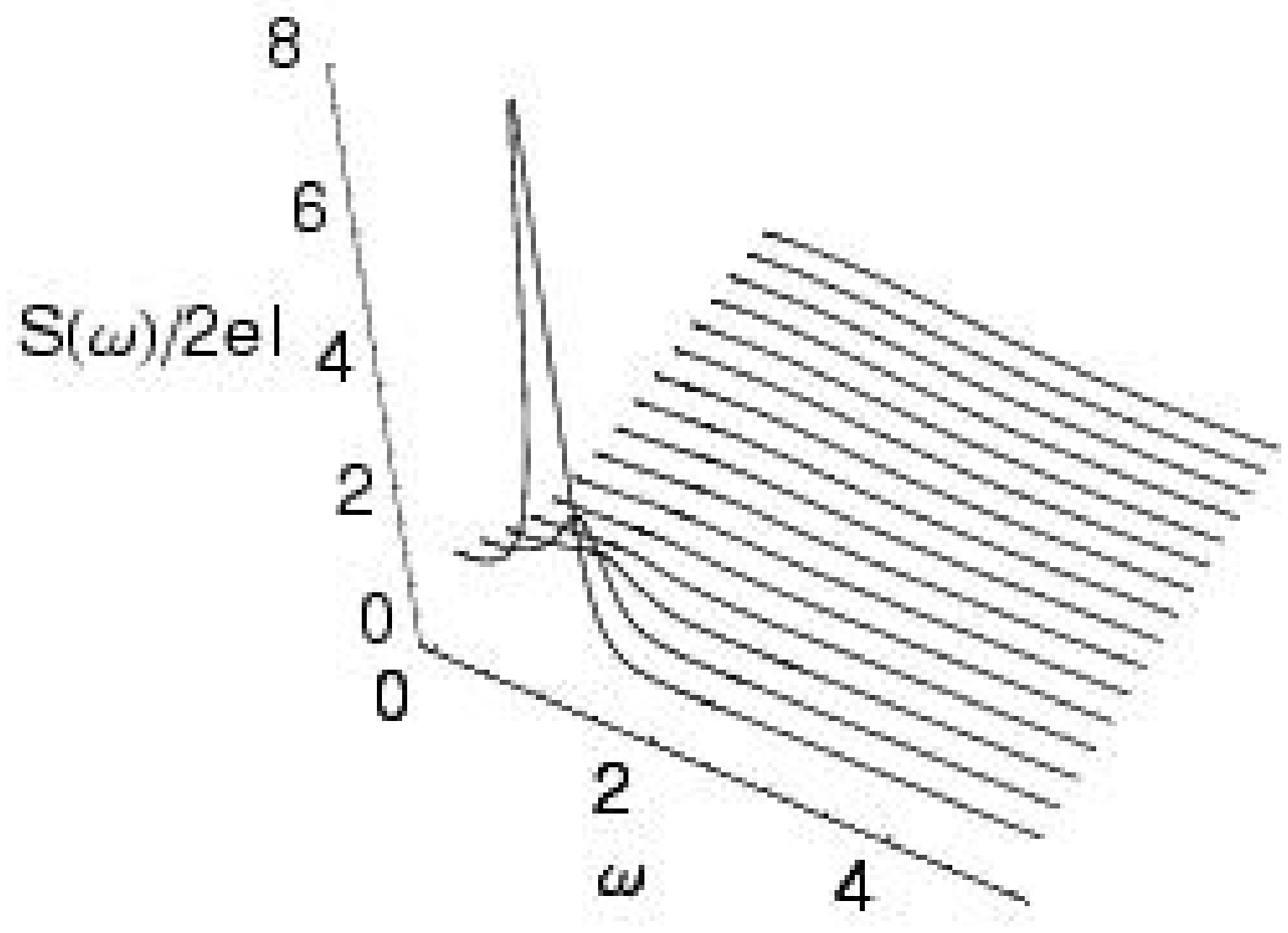}
\includegraphics[width=.45\linewidth]{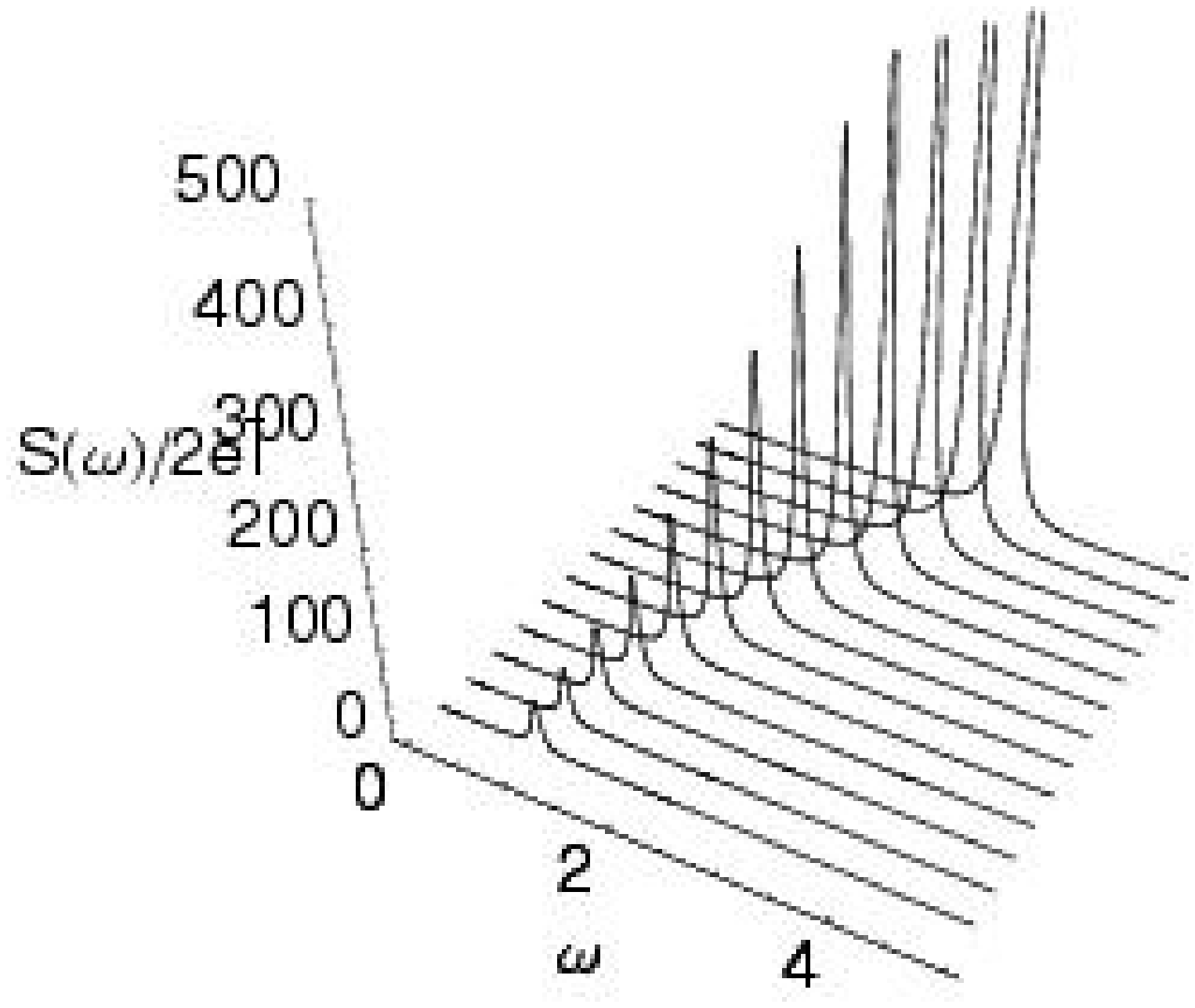}
\caption{Current noise power, (a) at resonance ($\veps=0$)
  for $\Gamma/E_J=0,.25,\cdots,5$, and (b) at a fixed weak dephasing
  rate ($\Gamma=0.1 E_J$) for $\veps/E_J=0,.25,\cdots,5$.}
\label{fig10}
\end{figure}

\begin{figure}
\includegraphics[width=.48\linewidth]{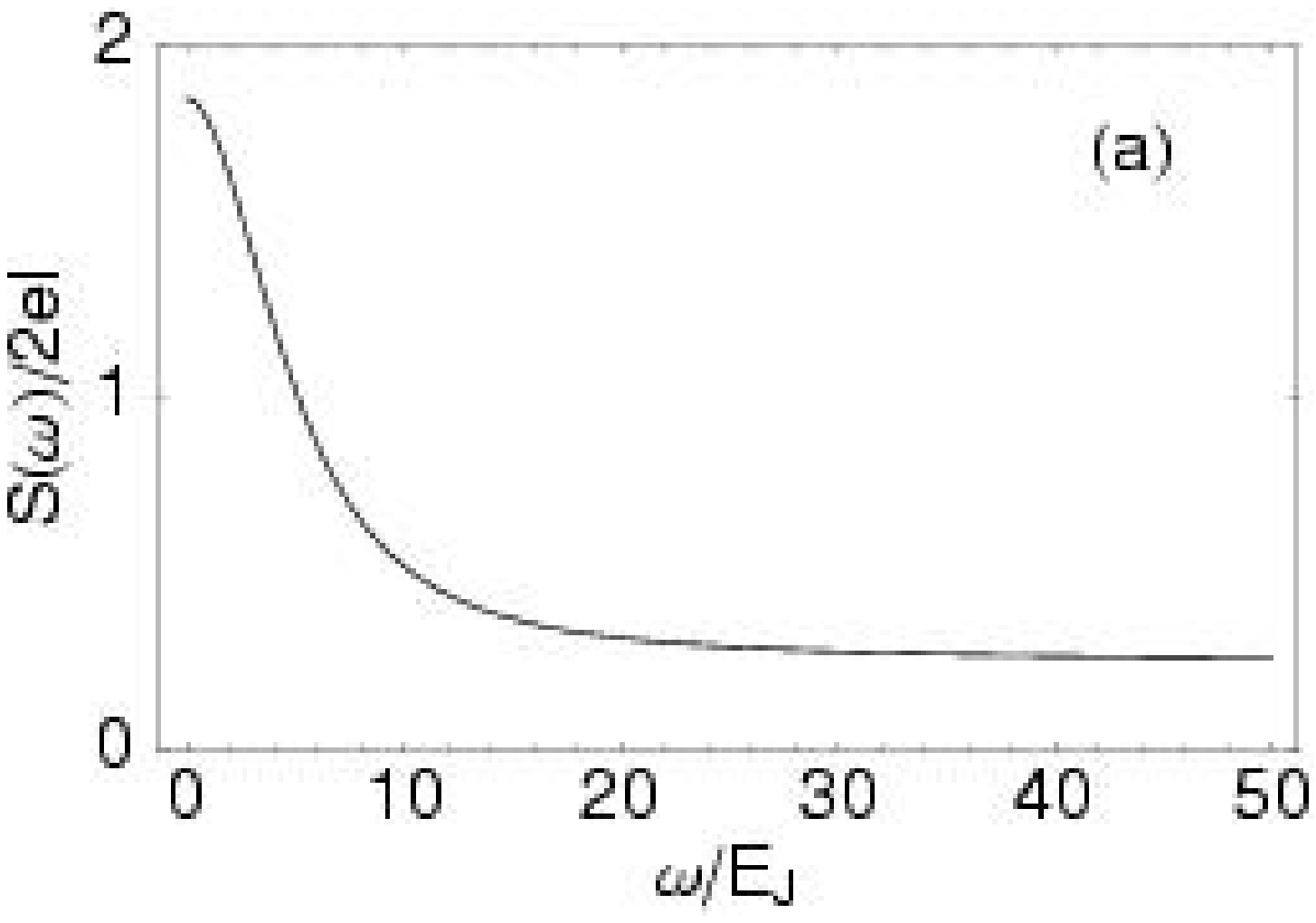}\
\includegraphics[width=.48\linewidth]{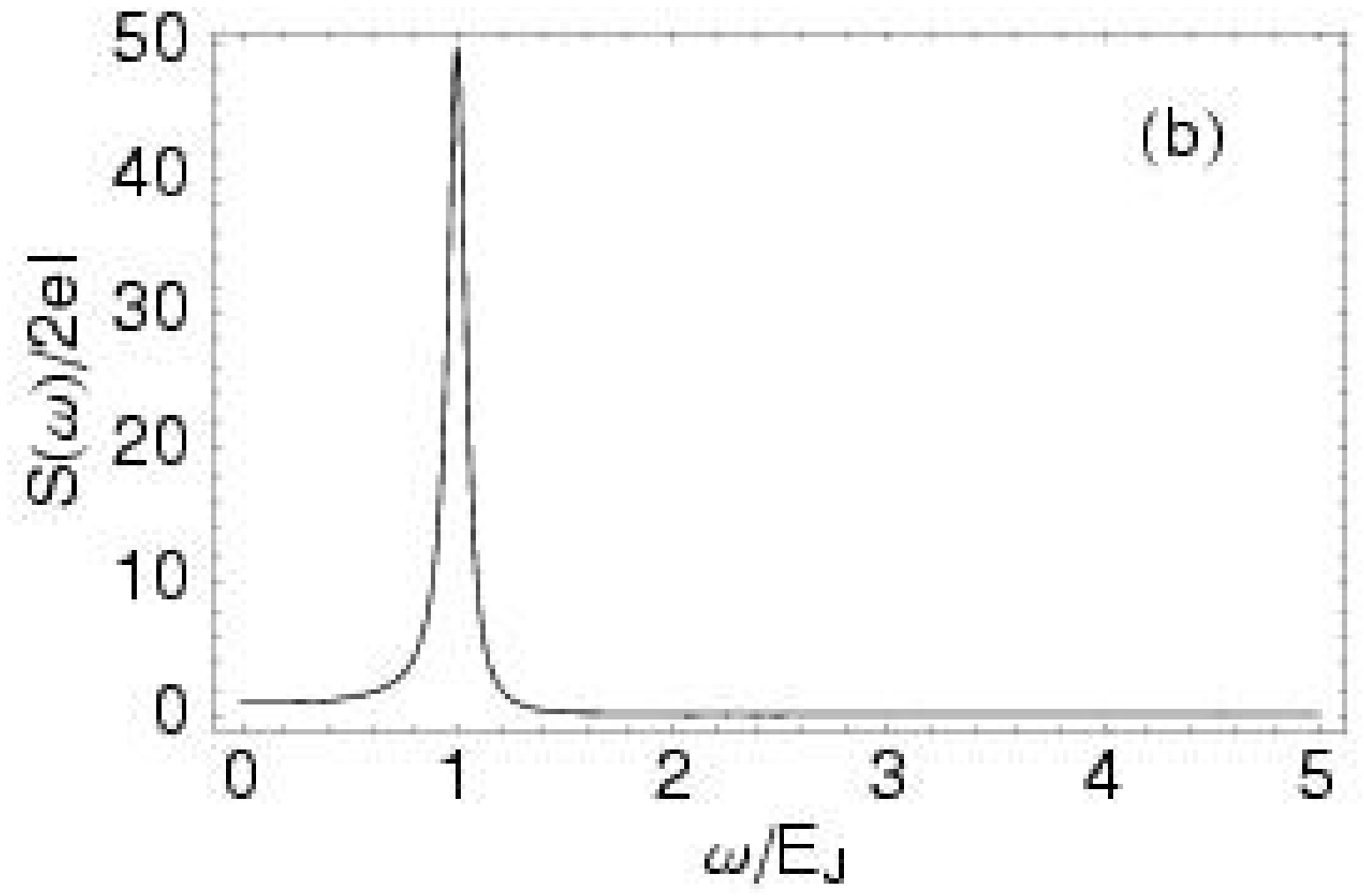}
\caption{Typical behavior of noise power spectrum $S(\omega)$ as a
  function of frequency $\omega$ in the (a) strong
  ($\Gamma_{1,2}\gg{E_J}$) and (b) weak ($\Gamma_{1,2}\ll{E_J}$)
  quasi-particle tunneling limits.  For both plots,
  $\veps=0$ and $C_L=C_R=C_\Sigma/2$ were assumed.}
\label{fig11}
\end{figure}

\subsection{Finite-Frequency Noise}

In Figs.~\ref{fig10} and \ref{fig11} we show the typical behavior of
the finite-frequency noise spectrum in the (a) strong and (b) weak
dephasing limits. It is interesting to notice that (only) in the weak
dephasing limit ($\Gamma\ll{E_J}$), there is a resonance peak of the
form
\begin{equation}
\label{noise:S4}
\frac{S(\omega)}{2eI} \approx \frac{C_R^2}{2C_\Sigma^2}\,
  \frac{E_J^2+2\veps^2}{(\omega-\omega_0)^2+\Gamma^2/4} \,,
\end{equation}
where the resonance frequency is given by
\begin{equation}
\omega_0 \simeq \sqrt{E_J^2+\veps^2}
  \left[1- \frac{\Gamma^2(E_J^2 + 3 \veps^2/2)}{4(E_J^2+\veps^2)^2}
  \right] \,.
\end{equation}
Clearly, the peak is an effect of coherent quantum oscillations
between the two energy levels separated by
$\omega_0\xrightarrow{\Gamma\to0}\sqrt{E_J^2+\veps^2}$, induced by the
Josephson effect across the left junction.  As expected, the resonance
peak is reduced in its height and broadened in its width with
increasing $\Gamma$.  On the contrary, as $\veps$ increases, the peak
gets sharper and the peak height increases quadratically with $\veps$.
However, this should not be confused with the zero-frequency case,
where $\veps$ effectively enhances the decoherence effects.  As
$\veps$ increases, the Josephson oscillation across the left junction
becomes faster, and there are less chances that it is interrupted by
the quasi-particle tunneling across the right junction.  This, in
turns, implies that the coherent oscillation is better defined and the
spectral component (especially $S_{LL}(\omega)$) at frequency
$\omega_0$ is highly enhanced.  For a vanishingly small
quasi-particle tunneling rate, $S_{LL}(\omega)/2 e I$ would approximately
become a delta--like function, centered at $\omega=\sqrt{E_J^2+\veps^2}$.
However, one should not be misled by this result, since the noise is always
proportional to the average current, which vanishes in this limit.

It is worth mentioning here on the relation between this result and
the description of the noise output from linear detector~\cite{Averin00}.
In the setup considered in this work, the right
electrode has the role of the detector of Cooper pair oscillations;
since the total current in the circuit is due to quasi-particle tunneling
(i.e. it is a {\it dissipative} current), the output signal may be considered
as classical. This has to be compared to the case of a detector measuring
the charge on the island. There, the back-action of the detector was essential
to produce an observable result. In the case of the current, instead,
the "detector" is intrinsically part of the system and it couples to the
observed quantity in an essentially non-linear way.

\section{Conclusions}
\label{long:sec6}

In this paper we considered properties of the distribution of the
transmitted charge in a superconducting SET tuned close to a Cooper
pair resonance. The dominant process to the transport in the regime
considered here, is the JQP cycle, a process in which coherent Cooper
pair oscillations are accompanied by (incoherent) quasi-particle tunneling.
The interplay between the coherence and the strong Coulomb blockade
manifests itself in various ways both in the counting statistics and in
the shot noise.
We found two distinct regimes characterized by different ratios of the
time scales for dephasing and relaxation, $\tau_r \ll \tau_\varphi$ in
the strong dephasing limit or $\tau_r \sim \tau_\varphi$ in the
opposite case of weak dephasing.

A generic feature of the counting statistics, valid in both the
regimes, is its even-odd asymmetry related to the fact that charge
transport is mediated by the Cooper pair tunneling. Other properties
are more pronounced in one of the two regimes. An example is the
dependence of $P_t(N,\tau)$ on the initial time $t$. This is clearly
visible in the strong dephasing limit while quickly lost in the weak
dephasing regime since, because of the strong Josephson energy, the
state changes significantly before quasi-particles have any time to be
produced. Another important point is that the counting statistics is
not Poissonian, due to the relevance of correlations between different
tunneling events. As a consequence the Fano factor is different from
the classical value.  The maximal suppression of the zero-frequency
shot noise is observed when the quasi-particle tunneling rate is
comparable to the frequency scale of the coherent Cooper pair
oscillations.

We finally investigated the shot noise at finite frequencies, which shows
a resonance peak at the Josephson oscillation frequency. This maximum
can be interpreted as an effect of coherent quantum transitions
between the two energy levels involved in the transport phenomena in
the device.

\acknowledgments%
We thank D.V. Averin, Y. Blanter, and J. Siewert for very useful
discussions. We acknowledge financial support from European Community
(IST-FET-SQUBIT) and INFM (PAIS-TIN).  M.-S.C.\ acknowledges the
support from the Swiss-Korean Outstanding Research Efforts Award
program.

%%%%%%%%%%%%%%%%%%%%%%%%%%%%%%%%%%%%%%%%%%%%%%%%%%%%%%%%
%%%%%%%%%%%%%%%%%%%%%%%%%%%%%%%%%%%%%%%%%%%%%%%%%%%%%%%%%%%%%%%%%%%


\begin{thebibliography}{10}
\bibitem{foot} The last term in each of
Eqs.~\eqref{long:C-RR}-\eqref{long:C-LR} describes the
auto-correlation of single current spikes and arises because the
tunneling events are considered to be instantaneous, see
Refs.~\onlinecite{Davies92,Hershfield93,Korotkov94a}.
\end{thebibliography}
\end{document}